# Calorimetric Evidence of Strong-Coupling Multiband Superconductivity in Fe(Te$_{0.57}$Se$_{0.43}$) Single Crystal

J. Hu[1], T.J. Liu[1], B. Qian[1], A. Rotaru[2], L. Spinu[2], and Z.Q. Mao[1]*

[1] Department of Physics and Engineering Physics, Tulane University, New Orleans, Louisiana, 70118, USA

[2] Advanced Materials Research Institute and Department of Physics, University of New Orleans, New Orleans, Louisiana 70148, USA

## Abstract

We have investigated the specific heat of optimally-doped iron chalcogenide superconductor Fe(Te$_{0.57}$Se$_{0.43}$) with a high-quality single crystal sample. The electronic specific heat $C_e$ of this sample has been successfully separated from the phonon contribution using the specific heat of a non-superconducting sample (Fe$_{0.90}$Cu$_{0.10}$)(Te$_{0.57}$Se$_{0.43}$) as a reference. The normal state Sommerfeld coefficient $\gamma_n$ of the superconducting sample is found to be ~ 26.6 mJ/mol K$^2$, indicating intermediate electronic correlation. The temperature dependence of $C_e$ in the superconducting state can be best fitted using a double-gap model with $2\Delta_s(0)/k_B T_c = 3.92$ and $2\Delta_l(0)/k_B T_c = 5.84$. The large gap magnitudes derived from fitting, as well as the large specific heat jump of $\Delta C_e(T_c)/\gamma_n T_c \sim 2.11$, indicate strong-coupling superconductivity. Furthermore, the magnetic field dependence of specific heat shows strong evidence for multiband superconductivity.

*zmao@tulane.edu

The discovery of layered iron pnictides[1-10] and iron chalcogenide[11-15] superconductors has ushered in a new age of high-temperature superconductivity. The iron chalcogenide $Fe_{1+y}(Te_{1-x}Se_x)$ is structurally the simplest of the Fe-based superconductors. Although its Fermi surface is similar to those of iron pnictides[16-17], the parent compound $Fe_{1+y}Te$ displays unique antiferromagnetic order with in-plane wave vector $(\pi,0)$ [18-19]. This contrasts with iron pnictide parent compounds which exhibit in-plane antiferromagnetic wave vector $(\pi,\pi)$ connecting the hole and electron pockets of the Fermi surface[20-21]. Aside from $(\pi,0)$ magnetic correlations[18-19], iron chalcogenide is also characterized by incommensurate itinerant magnetic fluctuations near $(\pi,\pi)$, which develop to spin resonance in the superconducting (SC) state for optimally doped samples[22-23]. The competition of these two magnetic correlations leads to an unusual phase diagram for $Fe_{1+y}(Te_{1-x}Se_x)$, i.e. an intermediate phase with charge carrier localization occurs between the long-range antiferromagnetic state ($x < 0.09$) and bulk SC phase ($x > 0.29$)[24]. These unique characteristics make iron chalcogenide $Fe_{1+y}(Te_{1-x}Se_x)$ a model system for studying the physics of Fe-based superconductivity.

In this work, we have investigated SC properties of the optimally doped iron chalcogenide $Fe(Te_{0.57}Se_{0.43})$ with onset $T_c = 14.7$ K through specific heat measurements on high quality single crystal samples. Specific heat can provide critical information on the thermodynamic properties of the SC state; it is a powerful technique for probing low-energy quasi-particle excitations and has been extensively applied to study iron pnictide superconductors. A variety of intriguing properties of iron pnictide superconductors have been revealed from specific heat measurements. For instance, the electronic specific heat $C_e$ is found to not vanish in zero-temperature limit even for optimally-doped samples[25-34], indicating the presence of unpaired quasi-particles. For the specific heat jump $\Delta C(T_c)$ near $T_c$, a unique power-law scaling behavior, i.e. $\Delta C(T_c) \sim T_c^3$, is observed[35]. This scaling law cannot be understood in term of BCS-theory, but implies that quantum criticality may play an important role in mediating superconductivity[36]. Critical information on superconducting pairing symmetry is also revealed from specific heat measurements. The electronic specific heat at SC state can be described well using a two-band model with isotropic *s*-wave gaps[28-31] or anisotropic *s*-wave gaps[34, 37].

Given the uniqueness of iron chalcogenide superconductors, it would be of particular interest to clarify if the specific heat of iron chalcogenide superconductors exhibit properties similar to those of iron pnictide superconductors. The analysis of specific heat data, however, is not straightforward, since this material possesses a very high upper critical field[38-41], as do the iron pnictide superconductors[42-46], which makes it difficult to separate the electronic specific heat from the phonon contribution through measurements on the normal state achieved by applying a magnetic field. Current approaches for the evaluation of SC electronic specific heat of iron pnictides is based on theoretical fitting of normal state specific heat data[25, 32, 47], or using a non-SC reference sample to evaluate phonon specific heat[26, 28-31, 33-34]. There have been a few reports on specific heat measurements on optimally doped sample $Fe_{1+y}(Te_{1-x}Se_x)$ with $x \sim$ 0.4-0.5, in which the phonon contribution is estimated by the theoretical fitting of normal state specific heat[48-51]. In this work, we present specific heat data analyses of $Fe(Te_{0.57}Se_{0.43})$ by using a Cu-substituted non-SC $(Fe_{0.90}Cu_{0.10})(Te_{0.57}Se_{0.43})$ sample as a reference to extract electronic specific heat for the SC sample. The electronic specific heat obtained from our analyses reveal the nature of strong-coupling multiband superconductivity in iron chalcogenide superconductors.

The SC Fe(Te$_{0.57}$Se$_{0.43}$) and reference non-SC (Fe$_{0.90}$Cu$_{0.10}$)(Te$_{0.57}$Se$_{0.43}$) single crystals were synthesized using a flux method as reported before[52]. Both samples are shown to be in pure tetragonal phases with space group P4/nmm by X-ray diffraction. The compositions were analyzed using energy dispersive X-ray spectrometer, and the excess Fe is shown to be less than 1% in both samples. The DC magnetization was measured using superconducting quantum interference device (SQUID, Quantum Design) under magnetic field of 30 Oe with zero-field cooling. The resistivity and specific heat were measured using the four-probe method and the adiabatic thermal relaxation technique respectively, in the Physical Property Measurement System (Quantum Design).

Figure 1 presents DC susceptibility and in-plane resistivity data for both SC Fe(Te$_{0.57}$Se$_{0.43}$) and reference (Fe$_{0.90}$Cu$_{0.10}$)(Te$_{0.57}$Se$_{0.43}$) samples. The bulk superconductivity of the SC sample is manifested in the susceptibility, which exhibits a sharp diamagnetic transition at ~14 K and reaches full diamagnetic screening ($4\pi\chi \sim -1$) within 1.5 K (Fig. 1a). This is consistent with the resistivity data which shows a sharp SC transition at $T_c \sim$ 15K with the transition width less than 1 K (Fig. 1b). Conversely, the reference sample does not display any trace of superconductivity in either susceptibility or resistivity, but exhibits insulating-like behavior at low temperature in resistivity. Such insulating-like behavior induced by Cu doping is consistent with the earlier report on the Cu-doping effect on FeSe[53].

The specific heat data of the SC and reference samples are presented in Fig. 2, which shows that both samples have comparable specific heat at temperatures above $T_c$ of the SC sample, indicating that both samples have similar phonon contributions to the specific heat. Therefore the Cu-doped sample is indeed an ideal reference sample for evaluating the phonon specific heat of the SC sample. In the SC sample, we observe a remarkable SC anomaly peak in specific heat at ~14 K. The specific heat jump $\Delta C(T_c)/T_c$, estimated by the isoentropic construction (see the right inset to Fig. 2), is ~ 51.0 mJ/mol K$^2$; this value is much larger than those of pnictide superconductors with comparable $T_c$ (e.g. $\Delta C(T_c)/T_c \approx 12$ mJ/mol K$^2$ for Ba(Fe$_{1-x}$Co$_x$)$_2$As$_2$ and Ba(Fe$_{1-x}$Ni$_x$)$_2$As$_2$ with $T_c$ ~15 K[35]), and does not follow the power-law scaling behavior of $\Delta C(T_c) \sim T_c^3$ mentioned above. A similar observation was also reported by Klein et al.[50] At temperatures well below $T_c$, we find that the data can be well fitted to $C = \gamma_{res}T + \beta T^3$, as shown in the left inset of Fig. 2, where $\gamma_{res}T$ and $\beta T^3$ represent the residual electronic specific heat and the phonon specific heat respectively. This indicates that in our SC sample there also exist residual electrons/holes which do not form Cooper pairs in zero-temperature limit, as seen in iron pnictide superconductors[25-34]. The linear fitting of $C/T$ versus $T^2$ yields $\gamma_{res} \approx 2.3$ mJ/mol K$^2$ and $\beta \approx 0.60$ mJ/mol K$^4$; $\gamma_{res}$ reaches 8.6% of the normal state Sommerfeld coefficient $\gamma_n$ ($\approx 26.6$ mJ/mol K$^2$, see below), comparable to that of optimally-doped iron pnictide superconductors ($\gamma_{res}/\gamma_n \sim$ 6-20%)[25-34]. This behavior was reproduced in specific heat measurements of several other SC samples taken from the same batch. Although no consensus has been reached on whether such residual electronic specific heat is associated with nodes in the superconducting gap or the pair-breaking effect caused by disorders or impurities for pnictide superconductors, our analyses given below suggest that the residual electronic specific heat in optimally doped Fe(Te$_{1-x}$Se$_x$) most likely results from disorder-induced pair-breaking effect.

In contrast with the SC sample, the specific heat of the reference sample does not show any SC anomaly, but can be fitted to $C = \gamma T + \beta^{ref}T^3$ at low temperature with $\gamma = 14.3$ mJ/mol

K$^2$ and $\beta^{\mathrm{ref}}$ = 0.71 mJ/mol K$^4$, as shown in the left inset of Fig. 2. Thus the phonon specific heat of the reference sample $C_{\mathrm{ph}}^{\mathrm{ref}}$ can be obtained by subtracting the electronic contribution $\gamma T$ from the measured specific heat. Since the SC and reference samples share similar phonon specific heats as indicated above, the phonon specific heat of the SC sample can be evaluated using the specific heat of the reference sample. According to the corresponding state principle[54], it can be reasonably assumed that the phonon contributions to entropy $S_{\mathrm{ph}}$ for both the SC and reference samples follow the same reduced function $S_{\mathrm{ph}} = f(T/\theta)$, where $f$ is a universal function and $\theta$ is the material-dependent characteristic temperature. The phonon entropy of these two samples are thus related by $S_{\mathrm{ph}}^{\mathrm{SC}}(T) = S_{\mathrm{ph}}^{\mathrm{ref}}(rT)$, with $r$ being the weakly temperature dependent scaling factor. From the derivative of the phonon entropy, we obtain the relationship of phonon specific heat $C_{\mathrm{ph}}$ between the SC and reference samples, which can be expressed as $C_{\mathrm{ph}}^{\mathrm{SC}}(T) = \mathrm{A} \cdot C_{\mathrm{ph}}^{\mathrm{ref}}(\mathrm{B} \cdot T)$, where A and B are renormalization factors associated with the scaling factor $r$ and its derivative. Therefore the specific heat of SC sample can be represented by

$$C^{\mathrm{SC}}(T) = C_{\mathrm{e}}^{\mathrm{SC}}(T) + C_{\mathrm{ph}}^{\mathrm{SC}}(T) = C_{\mathrm{e}}^{\mathrm{SC}}(T) + \mathrm{A} \cdot C_{\mathrm{ph}}^{\mathrm{ref}}(\mathrm{B} \cdot T), \quad (1)$$

where $C_{\mathrm{e}}^{\mathrm{SC}}(T)$ is the electronic specific heat of the SC sample.

We fitted the normal state specific heat data of the SC sample in 25-40 K temperature range, where $C_{\mathrm{e}}^{\mathrm{SC}}(T) = \gamma_{\mathrm{n}} T$, using eq. (1) under the constraint of entropy conservation at onset $T_{\mathrm{c}}$ (i.e. $\int_0^{T_{\mathrm{c}}} C_{\mathrm{e}}^{\mathrm{SC}}(T)/T \mathrm{d}T = \int_0^{T_{\mathrm{c}}} \gamma_{\mathrm{n}} \mathrm{d}T$). Here the onset $T_{\mathrm{c}}$ for the entropy conservation constraint is ~14.7 K, which is determined from the sharp change of the derivative of $C^{SC}(T)/T$. The renormalization factors A and B derived from our fitting are 1.03 and 0.99 respectively; they are reasonably close to unity because of the similar phonon specific heat between the SC and reference samples. The $\gamma_n$ obtained from fitting is 26.6 mJ/mol K$^2$, in good agreement with those derived from photoemission spectroscopic measurements on FeTe$_{0.58}$Se$_{0.42}$ (29(6) mJ/mol K$^2$)[55], and earlier specific heat measurement on FeTe$_{0.5}$Se$_{0.5}$ (23-26 mJ/mol K$^2$) for which the phonon contribution at superconducting state was evaluated through the extrapolation of the normal state phonon specific heat[48-51]. This value of $\gamma_n$ is comparable with those of optimally Co-doped Ba(Fe$_{1-x}$Co$_x$)$_2$As$_2$ ($\gamma_n$ ~22 mJ/mol K$^2$)[26, 28-30, 33-34, 37], suggesting that the electronic correlation strength in iron chalcogenide superconductor is intermediate as in iron pnictide superconductors[56].

By subtracting the phonon contribution $A \cdot C_{ph}^{ref}(B \cdot T)$ from the measured specific heat of the SC sample, the electronic specific heat can be extracted as shown in Fig. 3. A pronounced jump at the SC transition is seen. The jump magnitude $\Delta C_e(T_c)/T_c$ is estimated to be ~56.0 mJ/mol K$^2$, slightly larger than that estimated directly from the measured specific heat (~ 51.0 mJ/K$^2$, see the right inset to Fig. 2). This difference can be attributed to the fact that the normal-state electronic specific heat $C_{en}/T$ gradually enhances as the temperatures approaches $T_c$ for $T$ < 25 K. This $C_{en}/T$ enhancement is not taken into account in the estimate of $\Delta C(T_c)/T_c$ shown in the right inset of Fig. 2. Similar $C_{en}/T$ enhancement near $T_c$ is also observed by other groups in specific heat measurements on similar samples [49-51]. There are two possible origins for such normal state electronic specific heat enhancement: SC fluctuations or magnetic spin fluctuations. Since no trace of superconductivity was probed above 16 K in any other measurements such as resistivity or susceptibility[12, 24, 57-59], SC fluctuations are less likely responsible for the observed $C_{en}/T$ enhancement. Spin fluctuations are therefore the most probable origin for the electronic specific heat enhancement. In fact, spin fluctuations in the normal state of iron chalcogenide superconductors have been observed in neutron scattering measurements[22-23, 60-61]. Additionally, NMR measurements on FeSe[62] show that normal state spin fluctuations enhance significantly as the temperature approaches $T_c$. Our observation of the $C_{en}/T$ enhancement near $T_c$ appears to imply a similar scenario for the optimally-doped Fe(Te$_{0.57}$Se$_{0.43}$). The reduced specific heat jump $\Delta C_e(T_c)/\gamma_n T_c$ evaluated from the electronic specific heat is 2.11, which is considerably larger than the BCS weak-coupling limit 1.43 and indicates strong-coupling superconductivity in iron chalcogenide superconductor. We note that this reduced specific heat jump is larger than that of optimally-doped Ba(Fe$_{1-x}$Co$_x$)$_2$As$_2$ ($\Delta C_e(T_c)/\gamma_n T_c \sim 1.5$)[26, 28-30, 33], but comparable to that of optimally-doped (Ba$_{1-x}$K$_x$)Fe$_2$As$_2$ ($\Delta C_e(T_c)/\gamma_n T_c \sim 2.5$)[25, 31, 47].

In addition to the large specific heat jump, the large SC energy gap derived from fitting of the temperature dependence of the electronic specific heat at superconducting state also supports the strong-coupling scenario. Figure 3 shows theoretical fits for the SC electronic specific heat. The dashed-line represents the fit based on the single band BCS $s$-wave model with an isotropic gap. All data points below $T_c^{onset}$ are nicely fitted within this model. The reduced gap value obtained from this fit, $2\Delta(0)/k_B T_c$, is about 5.18, much larger than the BCS weak-coupling limit $2\Delta(0)/k_B T_c = 3.53$, but consistent with the result reported in ref. 50 where $2\Delta(0)/k_B T_c = 5$. We note that a single-band fit with an isotropic gap was attempted in several other specific heat studies on samples similar to ours[48-49, 51]. The reduced gap reported in those studies ranges from 6.4[48] to 3.57[49, 51], more or less than the value from our single-band fit. This discrepancy most likely originates from the different estimates of phonon specific heat. As addressed above, the phonon specific heat of our SC sample is evaluated from the specific heat of the non-SC reference sample, whereas in previous studies the phonon specific heat of SC state is extrapolated from a theoretical fit of normal state specific heat, which can often lead to under- or overestimate.

As noted above, for iron pnictide superconductors, the electronic specific heat in the SC state can be described well using a two-band model with isotropic $s$-wave gaps[28-31] or anisotropic $s$-wave gaps[34, 37]. In order to examine whether this model works for iron chalcogenide superconductors, we have also tried fitting our data using the two-band model with isotropic gaps, presented by the solid-line in Fig. 3. As illustrated in the inset of Fig. 3, the two-band fit is improved over the single band fit at low temperature. This isotropic-gap fitting clearly suggests that the residual electronic specific heat observed in SC state ($\gamma_{res} \approx 2.3$ mJ/mol K$^2$) should be attributed to pair-breaking effect of disorders, rather than nodal gaps. In fact, disorders intrinsically exist in our sample since it is an alloy where disorders are unavoidable. The reduced gaps derived from the double-gap fit are $2\Delta_s(0)/k_BT_c = 3.92$ and $2\Delta_l(0)/k_BT_c = 5.84$, both larger than the BCS weak-coupling limit and the relative weight between the small and large gaps is ~ 0.57. The ratio of these two gaps, $\Delta_s(0)/\Delta_l(0)$, is ~0.7, which is noticeably larger than that seen in iron pnictide superconductors where $\Delta_s(0)/\Delta_l(0) = 0.3 \sim 0.5$ [28-31]. The large gap magnitudes derived from our fitting are clearly consistent with the strong-coupling superconductivity suggested by the large specific heat jump at $T_c$ described above. We note that strong-coupling superconductivity for iron chalcogenide has also been suggested by other experiments, though SC energy gaps probed in different experiments are not entirely consistent. Both point-contact Andreev reflection[63] and photoemission spectroscopy measurements[64] on samples similar to ours reveal a large single gap with $2\Delta(0)/k_BT_c = 6 \sim 7$. However, optical conductivity[65], $\mu$SR [66-67], and penetration depth[68] measurements suggest double gaps. Both gaps probed in optical conductivity experiments are larger than the BCS weak limit, with $2\Delta(0)/k_BT_c \approx 4.0$ and $8.4$[65], while the smaller gap revealed in $\mu$SR [66-67] and penetration depth[68] measurements is smaller than the BCS weak limit with $2\Delta_s(0)/k_BT_c = 0.8$ and $2.4$, $2\Delta_l(0)/k_BT_c = 4.0$ and $4.3$.

To add more insights to the pairing symmetry for iron chalcogenide superconductors, we have investigated the field dependence of specific heat for our SC sample. As shown in Fig.4a, the low temperature specific heat can be well fitted to $C/T = \gamma + \beta T^2$ for various magnetic fields. For fully gapped superconductors, the field induced quasi-particle density of states, represented by $\Delta\gamma(H) = \gamma(H) - \gamma(0)$, is expected to exhibit linear field dependence, since the quasi-particle states are proportional to the density of vortex cores which is linearly dependent on the magnetic field. Although the temperature dependence of SC specific heat of Fe(Te$_{0.57}$Se$_{0.43}$) can be fitted using the isotropic $s$-wave model, the field induced change in electronic specific heat $\Delta\gamma(H)$ does not follow typical isotropic $s$-wave behavior. Instead, $\Delta\gamma(H)$ is even lower than that expected for isotropic $s$-wave pairing (Fig. 4b). Such abnormal behavior is also reflected in the specific heat data reported recently by Klein *et al.* in Ref. 50 where measurements were conducted up to 28 T. We have included their high-field data in Fig. 4b for comparison. Apparently our $\Delta\gamma(H)$ data are quite consistent with their high field data; both follow a similar field dependence and lie below the linear line anticipated for isotropic $s$-wave paring. This unusual field dependence of $\Delta\gamma(H)$ is distinct from the behaviors observed in iron pnictide superconductors where $\Delta\gamma(H)$ exhibits either a linear field dependence[25, 27], or a sub-linear field dependence lying between the linear line expected for the *s*-wave and the $\sqrt{H}$ curve expected for the *d*-wave pairing [26, 28, 34, 37]. Since *d*-wave pairing has already been ruled out by a growing number of experiments, such as the observation of the *c*-axis Josephson effect[69] and the absence of the paramagnetic Meissner effect[70], the linear or sub-linear field dependence of $\Delta\gamma(H)$ in iron pnictides implies isotropic[25] or anisotropic *s*-wave pairing[28, 34, 37], or a multiband effect[26-27].

What is the origin of the slow increase of $\Delta\gamma(H)$ in iron chalcogenide superconductor? According to a recent theory, this is most likely associated with a multiband effect[71]. This theory indicates that for multiband superconductors with disorder/impurity scatterings, if the ratio of two isotropic *s*-wave gaps $\Delta_s/\Delta_l > 0.5$, the field-induced low energy excitation would be less remarkable compared to the single-band *s*-wave pairing. In this scenario $\Delta\gamma(H)$ would slowly increase with field for the low field region, but superlinearly rises to $\gamma_n$ at fields close to the upper critical field $H_{c2}$. As stated above, our double-gap fitting in Fig. 3 has revealed the ratio of $\Delta_s/\Delta_l$ to be ~ 0.7, which is indeed above the critical value of 0.5 suggested by the theory. As a result, our observation of slow increase in $\Delta\gamma(H)$ can be viewed as an evidence of multiband superconductivity for iron chalcogenide superconductors. The effect of disorder scattering on $\Delta\gamma(H)$ is also examined in Ref. 71. For the SC state with a sign-change order parameter, which is believed to be the case for Fe-based superconductors, disorder scattering-induced unpaired states near Fermi level would lead $\Delta\gamma(H)$ to be more sublinear in the low field region as the $\Delta_s/\Delta_l$ ratio is less than 0.5. This effect can be used to interpret the sublinear field dependence of $\Delta\gamma(H)$ in pnictide superconductors where $\Delta_s(0)/\Delta_l(0) = 0.3 \sim 0.5$ [28-31], but is not reflected in our data presented in Fig. 4 since in our sample $\Delta_s / \Delta_l$ is ~0.7, conspicuously greater than those of pnictide superconductors.

Finally, it is worthwhile to point out that although the fit for temperature dependence of SC electronic specific heat presented above suggests isotropic gaps, the actual pairing symmetry of iron chalcogenide superconductors may be far more complex. A widely-discussed multiband model predicts that gaps on hole bands are fully gapped, while electron bands have nodeless anisotropic gaps or nodal gaps[72-75]. We note that the recent angle-resolved low-temperature specific heat measurements on FeTe$_{0.55}$Se$_{0.45}$ reveal a remarkable four-fold oscillation of the specific heat with the in-plane rotation of magnetic field, which provides a strong support for gap anisotropy on the electron pockets[76].

In summary, we have investigated the temperature and field dependence of specific heat of optimally-doped iron chalcogenide superconductor Fe(Te$_{0.57}$Se$_{0.43}$). Using the specific heat of a non-SC sample (Fe$_{0.90}$Cu$_{0.10}$)(Te$_{0.57}$Se$_{0.43}$) as a reference has enabled us to separate the electronic specific heat from the phonon contribution for the SC sample. The nature of strong-coupling superconductivity is revealed from the large superconducting energy gap and the large specific heat jump near $T_c$. Our analyses also show that, although the electronic specific heat of superconducting state can be fitted using either a single-band or a two-band model with isotropic gaps, the change of electronic specific heat induced by magnetic field can be understood only in terms of multiband superconductivity. Disorders play an essential role in this superconductor; the pair-breaking caused by disorder-scattering should be responsible for the non-vanishing electronic specific heat in the zero-temperature limit. In addition, the normal state electronic specific heat coefficient derived from our analyses suggests intermediate electronic correlation in iron chalcogenides.

The work at Tulane is supported by the NSF under grant DMR-0645305 for materials and equipment, the DOE under DE-FG02-07ER46358 for personnel. Work at AMRI was supported by DARPA through Grant HR 0011-09-1-0047. The authors are grateful to Dr. Tanatar for informative discussions and Dr. Fobes for technical support.

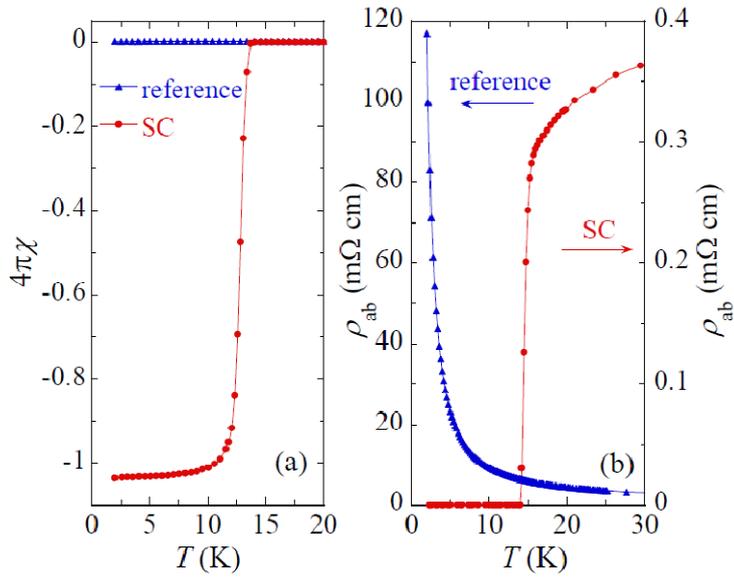

Figure 1: (Color online) (a) Temperature dependence of DC susceptibility measured with a magnetic field of 30 Oe (applied along the c-axis) and zero-field-cooling history; (b) In-plane resistivity as a function of temperature. SC and reference represent $Fe(Te_{0.57}Se_{0.43})$ and $(Fe_{0.90}Cu_{0.10})(Te_{0.57}Se_{0.43})$ samples respectively.

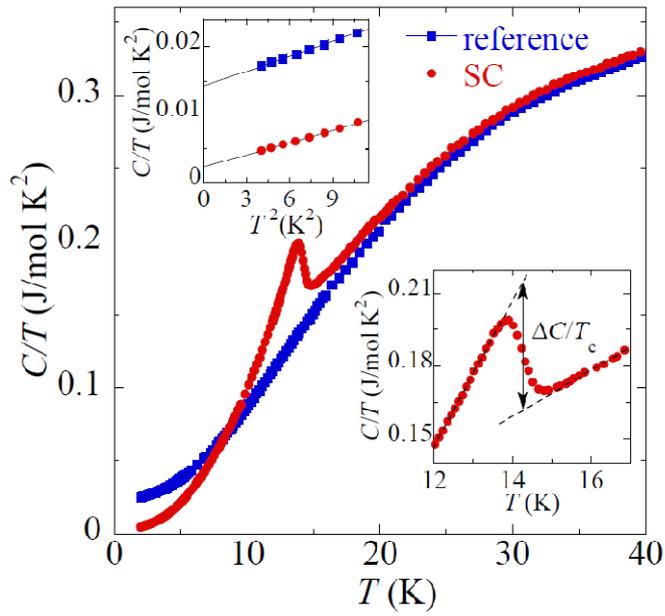

Figure 2: (Color online) Temperature dependence of specific heat $C(T)/T$ for SC and reference samples. Left inset: $C(T)/T$ vs. $T^2$ at low temperature for both samples. The solid line shows the linear fit to $C(T)/T = \gamma + \beta T^2$. Right inset: Specific jump $\Delta C(T_c)/T_c$ evaluated from the isoentropic construction.

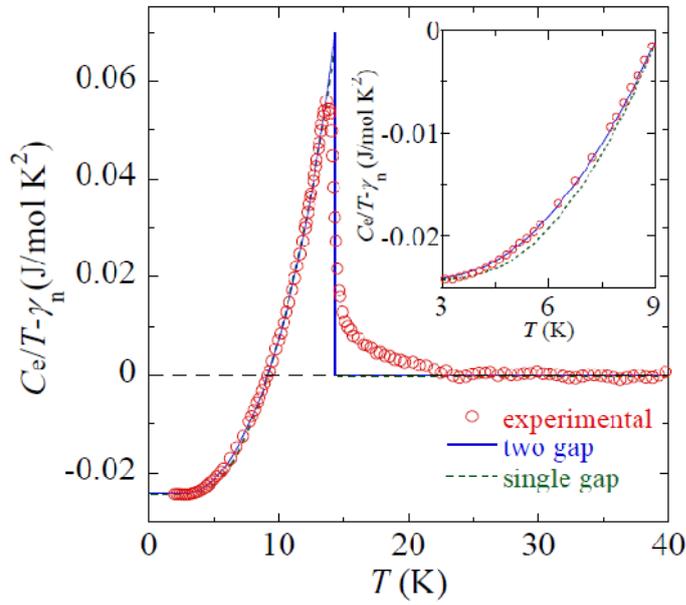

Figure 3: (Color online) Electronic specific heat $C_e(T)/T$ as a function of temperature for Fe(Te$_{0.57}$Se$_{0.43}$) (The data has been subtracted by the normal state Sommerfeld coefficient $\gamma_n$ ($\approx$ 26.6 mJ/mol K$^2$, see the text). The solid and dash lines represent the phenomenological two-band model fit and the BCS single-band model fit respectively. The inset shows a difference between the single- and two-band fits.

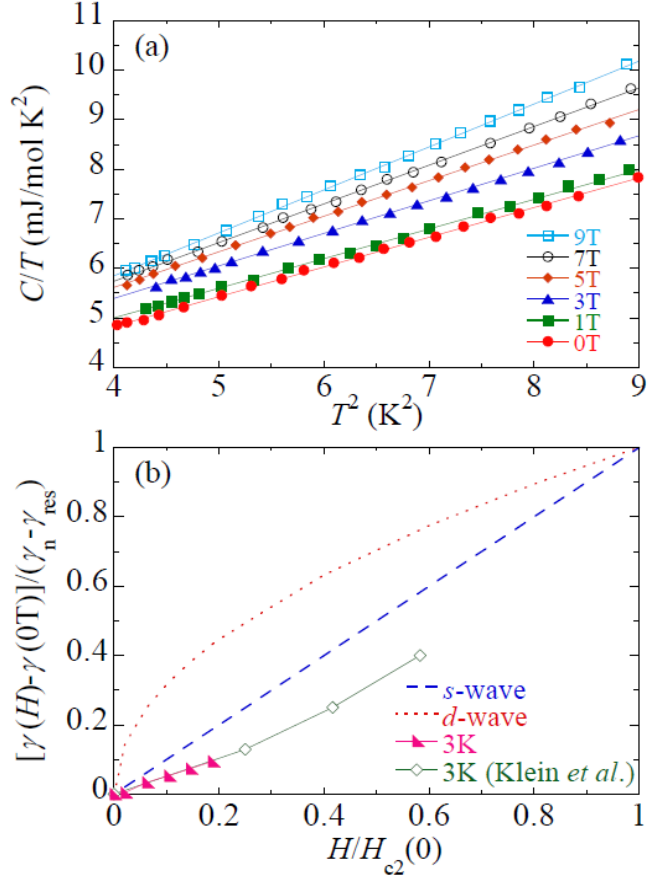

Figure 4: (Color online) (a) Low temperature specific heat $C/T$ as a function of $T^2$ under various magnetic fields applied along the crystalline $c$-axis for Fe(Te$_{0.57}$Se$_{0.43}$). The solid lines represent the linear fit to $C/T = \gamma + \beta T^2$. (b) Magnetic field-induced change in the specific heat at 3.0 K, normalized to $\gamma_n - \gamma_{res}$ where $\gamma_n$ and $\gamma_{res}$, respectively, represent the normal state Sommerfeld coefficient ($\approx 26.6$ mJ/mol K$^2$) and the coefficient of residual electronic specific heat ($\approx 2.3$ mJ/mol K$^2$, see the text). $H/H_{c2}(0)$ represents the reduced field, where $H_{c2}(0) = 48$ T, quoted from Ref. 40. The diamond symbol represents the $\Delta\gamma(H)$ data reported by Klein $et\ al.$[50]. The dashed and doted lines represent the field dependence of $\Delta\gamma$ expected for the standard $s$-wave and the clean $d$-wave pairing respectively.